\documentclass[12pt]{article}
\usepackage[latin1]{inputenc}
\usepackage{pstricks}
\usepackage{amsmath}
\usepackage{float}
\usepackage{color}
\input epsf
\usepackage{amssymb}
\usepackage[dvips]{graphicx,psfrag}
\newcommand{\be}{\begin{equation}}
\newcommand{\ee}{\end{equation}}
\newcommand{\bea}{\begin{eqnarray}}
\newcommand{\eea}{\end{eqnarray}}
\newcommand{\lb}{\label}
\begin{document}
\begin{titlepage}
\title{Dispersion in the growth of matter perturbations}
\author{R. Gannouji$^a$\thanks{email:gannouji@iucaa.ernet.in}~,
B. Moraes$^b$\thanks{email:moraes@lpta.univ-montp2.fr}~and
D. Polarski$^b$\thanks{email:polarski@lpta.univ-montp2.fr}\\
\hfill\\
$^a$~IUCAA, Post Bag 4, Ganeshkhind, Pune 411 007, India\\
\hfill\\
$^b$~Lab. de Physique Th\'eorique et Astroparticules, CNRS\\ 
Universit\'e Montpellier II, France}
\pagestyle{plain}
\date{\today}

\maketitle

\begin{abstract}
We consider the linear growth of matter perturbations on low redshifts 
in modified gravity Dark Energy (DE) models where $G_{\rm eff}(z,k)$ is 
explicitly scale-dependent. 
Dispersion in the growth today will only appear for scales of the 
order the critical scale $\sim \lambda_{c,0}$, the range of the 
fifth-force today.
We generalize the constraint equation satisfied by the parameters 
$\gamma_0(k)$ and $\gamma'_0(k)\equiv \frac{d\gamma(z,k)}{dz}(z=0)$ 
to models with $G_{{\rm eff},0}(k)\ne G$. Measurement of $\gamma_0(k)$ 
and $\gamma'_0(k)$ on several scales can provide information about 
$\lambda_{c,0}$. 
In the absence of dispersion when $\lambda_{c,0}$ is large compared 
to the probed scales, measurement of $\gamma_0$ and $\gamma'_0$ 
provides a consistency check independent of $\lambda_{c,0}$. This 
applies in particular to results obtained earlier for a viable $f(R)$ 
model. 

\end{abstract}

\end{titlepage}

\section{Introduction}
The present accelerated expansion of the universe \cite{P97}is a major 
problem facing cosmologists \cite{SS00}. If it is due to some smooth 
(isotropic) perfect fluid, its 
pressure should be sufficiently negative in order to induce an accelerated 
expansion. The simplest candidate, and maybe even the oldest one, is a 
cosmological constant $\Lambda$. However its required magnitude can be 
viewed as unnaturally small in order to provide a satisfactory solution 
of the puzzle. It is then natural to look for other alternative models 
providing an effective cosmological constant in the late-time universe 
whose origin is dynamical. A more drastic step is to assign the present 
accelerated expansion to a modification of gravity on cosmic scales. 

It was realized some time ago that the simultaneous study of the background 
expansion and of the matter perturbations growth could provide a mean 
to break the degeneracy between DE models based on different gravity 
theories \cite{S98}, (see also e.g.\cite{B07}). For this reason the study of the matter 
perturbations growth has been the subject of many investigations in recent 
years. An important goal is to find appropriate characterizations of the 
perturbations growth allowing to discriminate between DE models based on General 
Relativity and those outside GR  and a convenient such characterization is the 
$\gamma$ formalism in which one writes the growth factor $f$ as 
$f=\Omega_m^{\gamma}$ \cite{HL06}. 

It was found that in models outside General Relativity, even on very low 
redshifts $\gamma$ must be treated as a non-constant function \cite{PG08}. 
In particular, it is possible to have a large slope 
$\gamma'_0\equiv \frac{d\gamma}{dz}(z=0)$
in some scalar-tensor models \cite{GP08} and $f(R)$ models \cite{GMP08} while 
this is not the case for standard DE models inside GR \cite{PG08}. 
An additional interesting phenomenon occuring in some modified gravity 
models is the appearance of dispersion in the growth of matter perturbations 
even for scales well inside the Hubble radius. Of course such a dispersion 
is absent in all DE models inside GR. It is then important to investigate 
if and how this dispersion could manisfest itself in the function $\gamma$ 
on very low redshifts. This will the focus of this letter. 

As we will show there is a straightforward way to check for the presence of 
dispersion and this method can be used to probe the parameter space of 
a given model to find which values will allow for dispersion. 
Precise measurements of the behaviour of $\gamma(z,k)$ at small redshifts 
could then not only detect a departure from $\Lambda$CDM but also directly 
constrain the models' parameters and the kind of modified gravity model 
under consideration.
While we will use a fiducial model for the calculation of the matter perturbations 
growth, this model is just used for illustrative purposes. Our results 
will apply to a large class of modified gravity DE models for which the 
growth of matter perturbations exhibit a scale-dependence. 
 

\section{Modified gravity DE models}

We describe first how modified gravity DE models can affect the 
growth of matter perturbations. We consider spatially flat 
Friedman-Lema\^{\i}tre-Robertson-Walker (FLRW) universes with a time-dependent 
scale factor $a(t)$ and a metric 
\be
ds^{2}= g_{\mu\nu}~dx^{\mu} ~dx^{\nu} = -dt^{2}+a^{2}(t)~d{\bf x}^{2}~.
\ee
In many modified gravity DE models matter perturbations satisfy a modified 
equation of the type 
\be
\ddot{\delta}_m + 2~H~\dot{\delta}_m - 
                         \frac{3}{2}~\Omega_m~H^2~\frac{G_{\rm eff}(t,k)}{G}~\delta_m = 0~,\lb{del}
\ee
where the quantity $G_{\rm eff}(z,k)$ is given by the expression 
\be
G_{\rm eff}(z,k) = G \left( 1+ 2 \beta^2 \frac{ \frac{k^2}{a^2 m^2} }
                                                { 1 + \frac{k^2}{a^2 m^2} } \right)~.\lb{Geff} 
\ee
The mass $m$ is varying with time and $am$ is typically decreasing very
rapidly with the expansion, the details of this decrease is
model-dependent. In some viable $f(R)$ models expression (\ref{Geff}) (with constant $G$) applies 
only in the high curvature regime for which $F(R)\equiv \frac{df}{dR}\approx 1$ \cite{S07},
\cite{HS07},\cite{GMP08} and $m$ is the scalaron mass introduced in \cite{S80} 
while $\beta=\frac{1}{\sqrt{6}}$.

In some DE models outside GR like scalar-tensor models, $G_{\rm eff}$ appearing 
in (\ref{del}) does not depend on $k$ \cite{BEPS00}. 
Equation (\ref{del}) is a generalization to models where the effective gravitational constant 
depends not only on time (or on $z$) but also on the perturbations scales. 
This is how dispersion can appear in their growth and it is important to 
investigate when it can appear on small redshifts. Note that the scalar-tensor
models considered in \cite{GP08} do {\emph not} correspond to expression
(\ref{Geff}) not even in the limit $k\to \infty$.

We will investigate equation (\ref{del}) using a chameleon model with action 
\be
\mathcal{S}~=~\int{d^4x}\sqrt{-g}\left(\frac{R}{16\pi G} 
             -\frac{1}{2} g^{\mu\nu} \partial_{\mu}\phi\ \partial_{\nu}\phi - V(\phi)\right) 
             + \mathcal{S}_m \left[ \Psi_m; A^2(\phi)~g_{\mu\nu} \right]~,
\ee 
 with an exponential potential of the type \cite{BBDKW04}
\be
V = M^4~ e^{(\frac{M}{\phi})^n}~. 
\ee
In our fiducial model the mass appearing in (\ref{Geff}) satisfies $m^2\equiv V_{,\phi\phi}$ so it is possible to change the mass scale $m$ by varying the parameter $n$ of the
potential. Hence the parameter $n$ can be used in order to tune the critical 
scale $\lambda_c$ ($\lambda_c$ depends also marginally on $\beta$). 
For a  conformal factor $A=e^{\beta \phi/M_{PL}}$, $\beta$ 
is a free parameter \cite{BBDKW04}.
If we assume that the (chameleon) field $\phi$ sits in the minimum of the 
(effective) potential from the early stages of the universe on, then we have 
$\frac{\phi}{M_{PL}}\ll 1$ during the subsequent evolution until today. 
As a result the background evolution is nothing else than that of $\Lambda$CDM. 
Moreover the conformal factor $A(\phi)\equiv  e^{\beta \phi/M_{PL}}$, with 
$M_{PL}^{-2}\equiv 8\pi G$, satisfies  $A=1$ to very high accuracy, so it 
will disappear from equations and does not have to be considered here.  

We insist that our aim is not to explore chameleon models here. This specific 
model is used in order to illustrate the various ways in which modified gravity 
DE models can depart from $\Lambda$CDM through the growth of matter perturbations 
on small redshifts including dispersion.

The physical meaning of $G_{\rm eff}$ is that the gravitational potential (per 
unit mass) in real space is modified through the presence of a fifth-force 
deriving from a Yukawa potential 
\be
V(r) = - \frac{G}{r}~\left( 1 + 2\beta^2 ~e^{- m r} \right) ~.\lb{Vr}
\ee
It is obvious that $\beta$ enters as the coupling constant of the fifth-force 
which has a range $L\sim  m^{-1}$. 
If we introduce the characteristic scale $\lambda_c$ 
\be
\lambda_c = \frac{2\pi}{ m }~,\lb{lcr} 
\ee
we have obviously 
\be
G_{\rm eff}(z,\lambda) = G \left( 1+ 2 \beta^2 \frac{ \frac{\lambda_c^2}{\lambda^2} }
                                                { 1 + \frac{\lambda_c^2}{\lambda^2} } \right)~,
\ee
where we have set $\lambda\equiv a\frac{2\pi}{k}$. We have the asymptotic regimes 
\bea
G_{\rm eff} &=& G( 1 + 2~\beta^2 )~~~~~~~~~~~~~~~~~~~~~~~~\lambda\ll \lambda_c ~,\lb{as1b}\\ 
&=& G~~~~~~~~~~~~~~~~~~~~~~~~~~~~~~~~~~~~~\lambda\gg \lambda_c~,\lb{as2b}
\eea 
Equation (\ref{del}) can be recast in the form 
\be
\frac{df}{dN} + f^2 + \frac{1}{2} ( 1 - 3 w_{\rm eff}) f = 
                         \frac{3}{2} \frac{G_{\rm eff}}{G}~\Omega_m~.\lb{fw}
\ee
where $w_{\rm eff}\equiv  - 1 - \frac{2\dot{H}}{3H^{2}}$. In (\ref{fw}) we have 
defined $f=\frac{d \ln \delta}{d \ln a}$ and  $N\equiv \ln a$. 
A well-known way to describe the growth of perturbations is through $\gamma$ 
defined as follows $f=\Omega_m(z)^{\gamma(z)}$. For a modified equation of the 
type (\ref{del}) or (\ref{fw}), f can be also scale dependent so that one must 
write
\be
f(z,k) = \Omega_m(z)^{\gamma(z,k)}~.\lb{gammazk}
\ee 
The analysis we perform here applies to many models. What will change is the 
microscopic origin of the quantities $\beta$ 
and $\lambda_c$ and the background expansion which does not have to 
be exactly that of $\Lambda$CDM. 
The parameter $\beta$ can be used to tune the change of 
$G_{\rm eff}$ with respect to its value $G$ in GR. 
A gratifying property of our fiducial model is that for all model parameters 
values that we will consider the background evolution is that of 
$\Lambda$CDM so that it is completely fixed once $\Omega_{m,0}$ is known.  
In this way we can straightforwardly relate $\gamma(z,k)$ to the behaviour of 
$G_{\rm eff}(z,k)$ and compare with the growth in $\Lambda$CDM. 

\section{Growth of matter perturbations} 

As we will see some models can exhibit 
a dispersion in the growth of matter perturbations on low redshifts. This 
dispersion will appear in the function $\gamma(z,k)$. 
All behaviours can be understood from the form of $G_{\rm eff}(z,k)$. 
It is clear from (\ref{fw}) that $\frac{3}{2} \frac{G_{\rm eff}}{G} \Omega_m$ 
is the source term driving the growth of $f$. 
By construction, all our fiducial models have the same background expansion as 
in $\Lambda$CDM. Therefore the function $\Omega_m(z)$ will be the same for all 
models with same $\Omega_{m,0}$ and differences in the functions 
$\gamma(z)$ are directly related to differences in $f$. 
More specifically, all changes can be immediately traced back to 
the behaviour of $\frac{G_{\rm eff}}{G}$, at any redshift, meaning that if 
it is in the transition regime at a given redshift, then there will be 
dispersion in the growth at that redshift.
As we have seen $\frac{G_{\rm eff}}{G}$ has the two asymptotic regimes 
(\ref{as1b})and (\ref{as2b}) and smaller scales are the first to 
undergo the transition from the large scales regime (\ref{as2b}) to the small 
scales regime (\ref{as1b}). 
The smaller the scale, the higher the transition redshift. 

We are interested in some interval of cosmic scales 
\be
\lambda_{\rm min}\le a_0 \frac{2\pi}{k} \le \lambda_{\rm max}~.\lb{int}
\ee
We will often illustrate our results with the concrete values 
\bea
\lambda_{\rm min} &=& 1 h^{-1}{\rm Mpc}\\ 
\lambda_{\rm max} &=& 30 h^{-1}{\rm Mpc}~.
\eea
Three main situations can arise depending on the critical length $\lambda_{c,0}$ today. 

\subsection{Small critical length $\lambda_{c,0}$}

All cosmic scales in the interval (\ref{int}) satisfy 
\be
\frac{G_{\rm eff}(z=0,k)}{G}\equiv \frac{G_{{\rm eff},0}(k)}{G} = 1~,\lb{G1}
\ee
in other words they are still today in the large scales regime (\ref{as2b}).
This will hold for any model for which the critical length 
$\lambda_{c,0}$ is sufficiently small 
\be
\lambda_{c,0}\ll 1 h^{-1}{\rm Mpc}~,
\ee
or equivalently when the mass $m_0$ is sufficiently large that no modification 
of gravity is felt on these scales.
Then the function $\gamma(z)$ will be the same in our fiducial model 
as in $\Lambda$CDM for all low redshifts where (\ref{G1}) holds, it is 
quasi-constant and depends weakly on $\Omega_{m,0}$. 
Then matter perturbations on these scales will not help to distinguish 
that model from a model inside GR with same background expansion and 
there is evidently no dispersion in the growth. 

\subsection{Large critical length $\lambda_{c,0}$}

All scales in the interval (\ref{int}) satisfy 
\be
\frac{G_{{\rm eff},0}(k)}{G}= 1 + 2\beta^2~.\lb{G2}
\ee
Physically this means that these scales had time to make the 
transition from the large-scales to the small-scales regime.
This will happen for any model where the critical length satisfies
\be
\lambda_{c,0}\gg 30 h^{-1}{\rm Mpc}~.
\ee
For these models the growth of matter perturbations will be affected on low 
redshifts allowing to distinguish them from the growth in GR. 
However the growth on low redshifts where (\ref{G2}) holds is the 
same for all scales, hence there is no dispersion in the 
growth and $\gamma(z)$ does not depend on $k$. 
Of course dispersion can appear on larger redshifts because different scales 
start their transition at different times. 
We will have the same parameter $\gamma_0$ for all scales in 
the interval (\ref{int}) and depending on the value of $2 \beta^2$ this value 
can still differ significantly from $0.55$ unless $2 \beta^2\ll 1$. 
This situation is illustrated with Figure 1.

\begin{figure}
\begin{center}
\includegraphics[scale=.75]{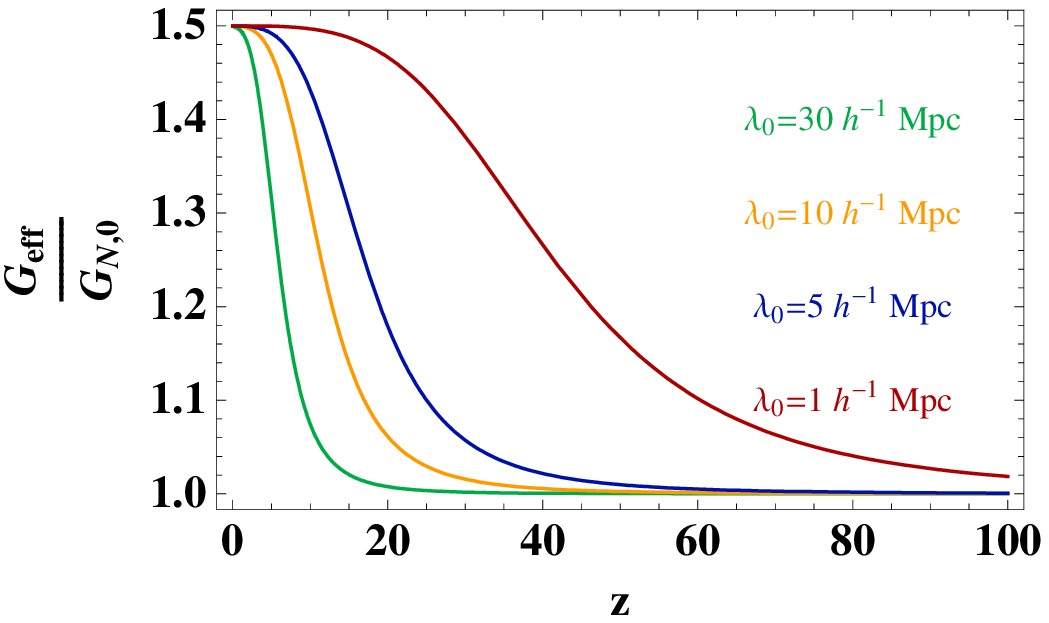} \includegraphics[scale=.7]{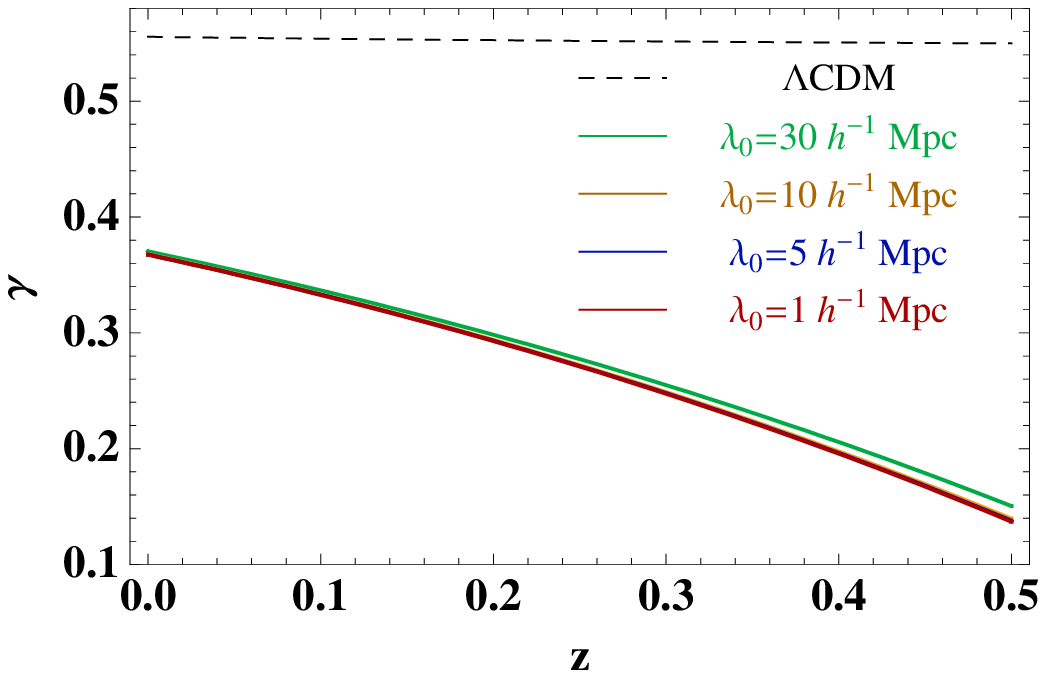}
\caption{ a) On the left panel, the behaviour of the quantity 
$\frac{G_{{\rm eff},0}(k)}{G}$ is shown for a very large critical length today, 
$\lambda_{c,0} = 1000 h^{-1}{\rm Mpc}$ and $\beta = 0.5$. The three smallest scales 
displayed have completed their transition to the small-scale regime (\ref{as1b}) on 
small redshifts while this is allmost the case for the largest scale 
$\lambda = 30 h^{-1}$Mpc. This 
is why a small dispersion can appear that distinguishes this scale from the other 
smaller scales. ~b) On the right panel, the corresponding functions $\gamma(z)$ are 
shown and a small dispersion is seen between the scale $\lambda = 30 h^{-1}$Mpc and 
smaller scales $\lambda \le 10 h^{-1}$Mpc. Of course, the model shown here is 
in trouble with observations because the value $\frac{G_{{\rm eff},0}(k)}{G}=1.5$ is 
rather high. Note also that $\gamma$ becomes negative in regions where 
$f>1$ while $\Omega_m(z)<1$, which is here the case on very low redshifts because of 
the high value of $G_{\rm eff}$ on these redshifts.}  
\end{center}
\end{figure}


\subsection{Intermediate critical length $\lambda_{c,0}$}

Finally new features appear if a given scale is today just in the transition 
between both regimes. For such scales we have 
\be
1 < \frac{G_{{\rm eff},0}(k)}{G} < 1 + 2 \beta^2~.\lb{G3}
\ee
This will happen in models for which 
\be
\lambda_0 \sim \lambda_{c,0}~.
\ee
The critical length should neither be much larger than $\lambda_{\rm max}$ 
nor much smaller than $\lambda_{\rm min}$. 
For these models we get a larger spectrum of behaviours as can be seen from 
Figure 2. 
We can have a pronounced dispersion in the growth of matter perturbations 
on small redshifts with scale-dependent $\gamma(z,k)$.  
Depending on whether $\frac{G_{\rm eff}(z=0,k)}{G}$ is close to one or not, large 
departures from the value 0.55 are obtained for $\gamma_0(k)$. When 
$\frac{G_{{\rm eff},0}(k)}{G}\approx 1$ we have a small departure, while for
$\frac{G_{{\rm eff},0}(k)}{G}\approx 1 + 2 \beta^2$ large departures can be obtained. 
Of course, if $2 \beta^2\ll 1$, the difference is essentially irrelevant. 
It could be in principle that the scales $\lambda_{\rm min}$ have 
completed their transition to the small-scales regime by today but not 
the larger scales. Higher (lower) values for $\gamma_0(k)$ are obtained 
for the larger (smaller) scales. 

\begin{figure}
\begin{center}
\includegraphics[scale=.75]{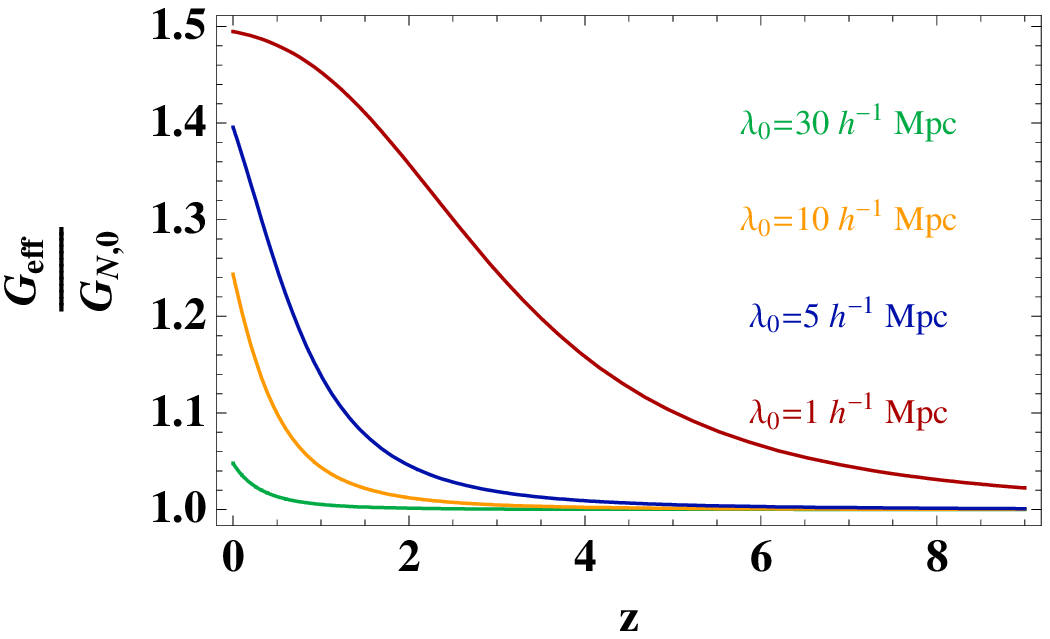} \includegraphics[scale=.7]{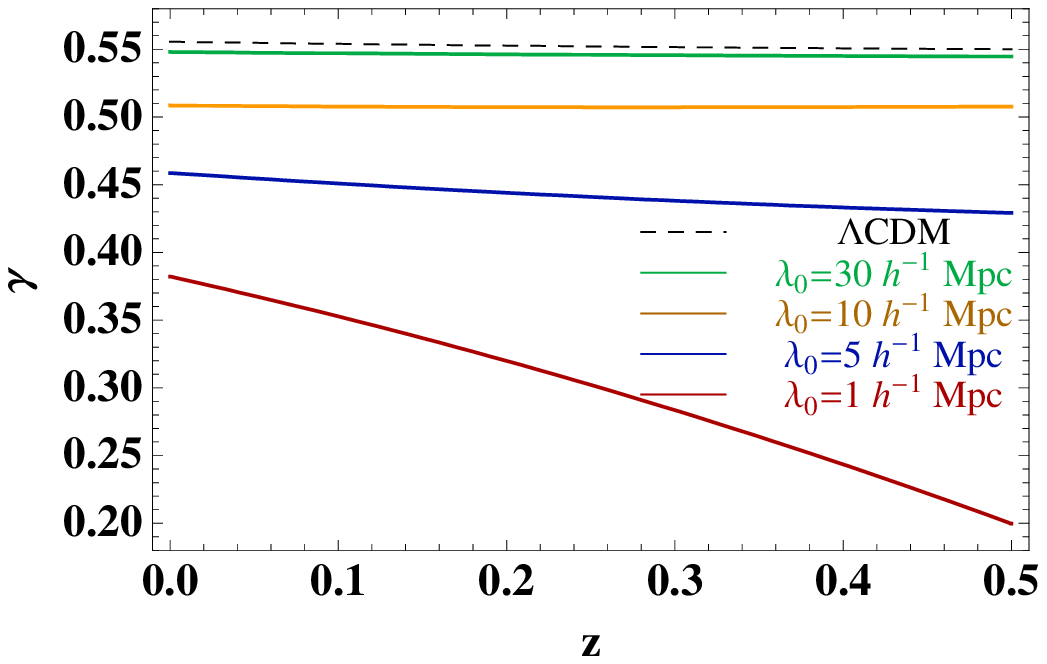}
\caption{ a) On the left panel, the behaviour of the quantity 
$\frac{G_{{\rm eff},0}(k)}{G}$ is shown when the critical length today 
$\lambda_{c,0}$ satisfies $\lambda_{c,0}=5 h^{-1}{\rm Mpc}$ while $\beta = 0.5$. Hence 
none of the scales displayed has completed its transition to the small-scale regime  
(\ref{as1b}). In particular, the value $\frac{G_{{\rm eff},0}(k)}{G}$ depends 
strongly on the scale and dispersion appears today in the growth of matter 
perturbations.~b) On the right panel, the corresponding functions $\gamma(z)$ are 
shown for which dispersion is evident. For large scales today $\lambda\gg \lambda_{c,0}$, 
$\gamma(z)$ is undistinguishable from its value in $\Lambda$CDM, but for 
$\lambda\lesssim \lambda_{c,0}$ it can be significantly different and large slopes 
$\gamma'_0$ are possible. Note in this respect that for our fiducial model, even 
large departures of $\gamma_0$ from $0.55$ can still yield a quasi-constant $\gamma(z)$ 
because this departure is conpensated by a value for $\frac{G_{{\rm eff},0}(k)}{G}$ 
significantly different from $1$ (see also Figure 3).} 
\end{center}
\end{figure}


\section{Important signatures and properties}

Let us consider all the characteristic signatures that can be obtained.
An important point concerns the behaviour of $\gamma(z,k)$. 
As emphasized in \cite{PG08}, while $\gamma(z)$ is quasi-constant on 
low redshifts for standard DE models inside General Relativity (GR), 
this is no longer true for DE models outside GR \cite{GP08},\cite{GMP08}. 
Actually there is a constraint equation which relates $\gamma_0(k)$ to the 
other background parameters. Indeed it is easy to derive the following 
equation which is valid for any $\gamma$ (we drop here the arguments for 
compactness)
\be
-(1+z)~\ln \Omega_m ~\gamma' + \Omega^{\gamma}_m +  \frac{1}{2}(1 + 3(2\gamma -1)~w_{\rm eff}) = 
                           \frac{3}{2}  ~\frac{G_{\rm eff}}{G}~\Omega^{1-\gamma}_m~.\lb{dgamma}
\ee 
From (\ref{dgamma}), it is easy to derive in turn the constraint equation 
\be
\gamma'_0 = \left[ \ln \Omega^{-1}_{m,0} \right]^{-1} ~\left[ -\Omega^{\gamma_0}_{m,0} - 
   3(\gamma_0 - \frac{1}{2})~w_{{\rm eff},0} + \frac{3}{2}~\frac{G_{{\rm eff},0}}{G} 
                                     \Omega^{1-\gamma_0}_{m,0} - \frac{1}{2}\right]~.\lb{dgamma0}
\ee
It is seen that the value of $\gamma'_0$ is fixed by the other parameters, we have a constraint 
equation (putting back the k dependence)
\be
C \left( \gamma_0(k),~\gamma'_0(k),~\Omega_{m,0},~w_{{\rm eff},0},
                ~\frac{G_{{\rm eff},0}(k)}{G} \right)=0~.\lb{C0}
\ee 
The constraint (\ref{C0}) generalizes the constraint found in \cite{PG08}  
when $\frac{G_{{\rm eff},0}}{G}=1$. In (\ref{C0}), $\frac{G_{{\rm eff},0}(k)}{G}$ 
is scale-dependent and can differ from one. 
We can parametrize this departure with a parameter $B$ defined as follows 
\be
\frac{G_{{\rm eff},0}(k)}{G} = 1 + 2B^2~. \lb{B}
\ee 
We have obviously $B^2\le \beta^2$, the equality is saturated for $\lambda_{c,0}\to \infty$, 
in practice this asymptotic regime is accurate up to $1 \%$ for scales satisfying 
$\lambda_c\gtrsim 10\lambda$. So (\ref{B}) can be used in two ways. Either it is used 
for scales that are in the small-scales regime (\ref{as1b}) and then we have $B=\beta$. 
Or it is used for scales in the intermediate regime and then $B^2$ measures the departure 
of $\frac{G_{{\rm eff},0}(k)}{G}$ from $1$ (GR), still different from $1+2\beta^2$. 
The constraint (\ref{C0}) is illustrated for representative background parameters in 
Figure 3.
\begin{figure}
\begin{center}
\includegraphics[scale=1]{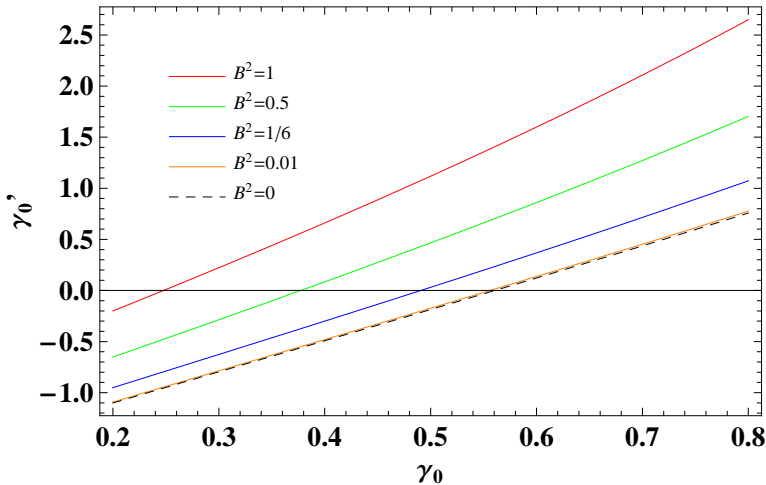} 
\caption{The quantity $\gamma'_0$ is shown as a function of  $\gamma_0$ for several 
values of $B^2$ defined in (\ref{B}) for the background parameters $\Omega_{m,0}=0.29$ 
and $w_{DE,0}=-1$. Note that $B^2\le \beta^2$, the equality holds in the small-scales 
asymptotic regime (\ref{as1b}). It is seen that as $B^2$ increases, quasi-static cases 
with $\gamma'_0\approx 0$ correspond to decreasing values of $\gamma_0$ significantly 
smaller than $0.55$. The value $B^2=0$ corresponds to General Relativity. Note further 
that $B^2=\frac{1}{6}$ corresponds to $f(R)$ models for scales in the small-scales 
asymptotic regime (\ref{as1b}).}  
\end{center}
\end{figure}

The actual occurence of the different possibilities must be addressed 
separately for each model and will be considered elsewhere but we can draw 
some general conclusions. The less interesting case is when $\lambda_{c,0}$ is 
so small that the growth on cosmic scales is like in GR. 
But in the other cases interesting signatures are obtained. 

If the background is known accurately we can hope to pinpoint the quantities 
$\Omega_{m,0}$ and $w_{{\rm eff},0}$ at the percent level.
If the characteristic length $\lambda_{c,0}$ which defines the range of the 
fifth-force today is large enough, no dispersion will be seen on scales 
(\ref{int}) if they satisfy $\lambda\ll \lambda_c$. Then the late-time growth 
on these scales takes place with a constant gravitational constant different 
from its value $G$ inside GR.
Nevertheless both $\gamma_0$ and $\gamma'_0$ can still depart strongly 
from the usual values inside GR if the coupling $\beta$ is large enough. 
This was indeed found for Starobinsky's $f(R)$ model \cite{GMP08}, remember 
there $1 + 2 \beta^2=\frac{4}{3}$.
For Starobinsky's model it was also found in \cite{GMP08} that  
$\lambda_{c,0}\sim H_0^{-1}$.
A large coupling $\beta$ will tend to lower the value of $\gamma_0$. 
For chameleon models, $\beta$ is a free parameter that can be recovered using 
(\ref{C0}) with  $\frac{G_{{\rm eff},0}(k)}{G}= 1+ 2 \beta^2$. 
For $f(R)$ models, $\beta$ is fixed but one can recover 
$F_0\equiv \frac{df}{dR}(z=0)$ using 
$\frac{G_{{\rm eff},0}(k)}{G} = \frac{4}{3 F_0}$ \cite{T07}. If we know $\Omega_{m,0}$ and $ w_{{\rm eff},0}$ we can find $F_0$ from the constraint (\ref{C0}). On the other hand if we know in addition $H_0$ we can also find independently $R_0$ and we can check whether $F_0 = F(R_0)$ if some $f(R)$ model is assumed. Hence simultaneous measurement of $\gamma_0$ and $\gamma'_0$ can serve as a consistency check in chameleon models if $\beta$ is known from other considerations or in $f(R)$ models if some given model is assumed. 

Finally in the intermediate case when $\lambda_{c,0}$ is of the order of the 
scales that are probed dispersion will appear in a way which depends on the exact 
location of $\lambda_{c,0}$ and it can be significant if $2 \beta^2$ is large. 
In the presence of dispersion, measurement of both $\gamma_0(k)$ and $\gamma'_0(k)$ 
on various scales can be used to reconstruct the quantity $G_{{\rm eff},0}(k)$. 
Such measurements can give information not only on $\beta$ but also on $\lambda_{c,0}$ 
as the constraint (\ref{C0}) must be satisfied for all scales in the transition regime.

To summarize, in a situation where there is no dispersion in $\gamma_0$ 
and $\gamma'_0$ these quantities provide important consistency checks. 
Even when dispersion does not affect the parameter  $\gamma_0$, it will 
still appear at large redshifts and induce a distortion of the power 
spectrum. On the other hand, a dispersion in the 
quantities $\gamma_0(k)$ and $\gamma'_0(k)$ could be a clear signature of 
the scale dependence of $G_{\rm eff}(k)$, and hence of a modification of gravity. 
It is however not clear whether it can appear in viable DE models. 

In any case, the growth of matter perturbations provides clearly a powerful 
discriminative tool in order to investigate whether or not the origin of the present 
accelerated expansion is due to a modification of gravity on cosmic scales.

\end{document}